# From Questions to Insights: Exploring XAI Challenges Reported on Stack Overflow Questions

Saumendu Roy  Saikat Mondal  Banani Roy  Chanchal Roy

University of Saskatchewan, Canada

{saumendu.roy,saikat.mondal,banani.roy,chanchal.roy}@usask.ca

## ABSTRACT

The lack of interpretability is a major barrier that limits the practical usage of AI models. Several eXplainable AI (XAI) techniques (e.g., SHAP, LIME) have been employed to interpret these models' performance. However, users often face challenges when leveraging these techniques in real-world scenarios and thus submit questions in technical Q&A forums like Stack Overflow (SO) to resolve these challenges. We conducted an exploratory study to expose these challenges, their severity, and features that can make XAI techniques more accessible and easier to use. Our contributions to this study are fourfold. First, we manually analyzed 663 SO questions that discussed challenges related to XAI techniques. Our careful investigation produced a catalog of seven challenges (e.g., disagreement issues). We then analyzed their prevalence and found that model integration and disagreement issues emerged as the most prevalent challenges. Second, we attempt to estimate the severity of each XAI challenge by determining the correlation between challenge types and answer metadata (e.g., the presence of accepted answers). Our analysis suggests that model integration issues is the most severe challenge. Third, we attempt to perceive the severity of these challenges based on practitioners' ability to use XAI techniques effectively in their work. Practitioners' responses suggest that disagreement issues most severely affect the use of XAI techniques. Fourth, we seek agreement from practitioners on improvements or features that could make XAI techniques more accessible and user-friendly. The majority of them suggest consistency in explanations and simplified integration. Our study findings might (a) help to enhance the accessibility and usability of XAI and (b) act as the initial benchmark that can inspire future research.

## CCS CONCEPTS

• **Software and its engineering** → *Software repository mining*; *Software maintenance and evolution*; *Practitioners' perspective*; *Challenges of XAI techniques.*

## KEYWORDS

Explainability Challenges, Stack Overflow, LIME, SHAP, Software Maintenance

## 1 INTRODUCTION

The growing importance of XAI lies in its ability to make AI (e.g., Machine Learning) models more transparent and understandable. This transparency builds trust and allows users to make informed decisions based on the insights provided by the model. Popular XAI techniques such as LIME (Local Interpretable Model-agnostic Explanations) [30] and SHAP (SHapley Additive Explanations) [16] have been widely adopted to explain model decisions. However, users often face challenges when applying these methods in real-world scenarios, and thus seek solutions by submitting questions to Technical Q&A forums like SO. Consider the example in Figure 1, where a user encounters difficulty applying SHAP's *TreeExplainer* to a scikit-learn *StackingClassifier*. In particular, the user receives an error that states the model type is not supported, which reflects a *compatibility issue* between XAI techniques (e.g., SHAP) and specific model types. However, this is a representative example of the many challenges users encounter when leveraging XAI techniques. Thus, it is crucial to identify all such challenges and address them to (a) improve the functionality of XAI tools, (b) enhance user accessibility, and (c) make AI systems more interpretable, trustworthy, and actionable.

Several studies have investigated the inconsistencies in post-hoc explanations [11, 28, 43]. For example, Roy et al. [35] and Parashar et al. [25] studied disagreements among post-hoc explanations in defect prediction models. Their studies suggested that different explainers could disagree in explaining a model, which challenged interpretability. Liao et al.[14] and Longo et al.[15] discussed the theoretical trade-off between performance and interpretability in black-box models. Liao et al. focused on how users understand explanations from these models, aiming to make them practical for decision-making. Meanwhile, Longo et al. explored design strategies for XAI that balance accuracy and ease of understanding in real-world contexts. These studies primarily focused on addressing specific challenges (e.g., disagreements among explainers), overarching problems (lack of trust and usability in explainability methods due to such disagreements) or exploring theoretical trade-offs. However, their studies did not uncover all the major practical challenges developers encounter when using XAI techniques in their tasks (e.g., installation, compatibility). Additionally, to the best of our knowledge, no existing study focuses on user-driven recommendations to improve the usability of XAI techniques.

```
Shap python Model type not yet supported by
TreeExplainer: class 'sklearn.ensemble._stacking
.StackingClassifier

I tried to use Shap (Tree Explainer) for
sklearn.ensemble._stacking.StackingClassifier

explainer = shap.TreeExplainer(clf)
shap_values = explainer.shap_values(x)
shap.initjs()
return shap.force_plot(explainer.expected_value[1], shap_values[1], x)

But I got an error: Model type not yet supported by TreeExplainer: <class
'sklearn.ensemble._stacking.StackingClassifier'>

How can I use shap force_plot for sklearn StackingClassifier ?

python   machine-learning   scikit-learn   ensemble-learning   shap
```

**Figure 1: An example of XAI compatibility issue [24].**

In this study, we conducted a manual analysis of SO questions that discussed the challenges associated with XAI techniques (e.g., SHAP, LIME) and produced a catalog of challenges. We assessed the severity of these challenges by analyzing their correlation with answer metadata (e.g., the presence of accepted answers) and by gathering insights from practitioners through surveys. Additionally, we captured practitioners' perspectives to provide actionable recommendations to improve the accessibility, usability, and trustworthiness of XAI tools. The insights from this study could bridge the gap between theoretical advancements and practical applications. In particular, we answered four research questions and thus made four contributions in this study.

**RQ1. What challenges do developers encounter while using XAI techniques, and which are more prevalent than others?** Understanding the challenges developers encounter with XAI is crucial for addressing practical barriers, improving tools, and creating tailored instructional resources to bridge the gap between theory and application. To catalog these challenges, we manually analyzed 663 XAI-related SO questions and spent a total of 160 person-hours. We also report the prevalence of each challenge to identify which challenges practitioners encounter most frequently.

**RQ2. Can we estimate the severity of each XAI challenge using answer metadata such as the presence of answers, including the acceptable ones?** Estimating the severity of XAI challenges correlating with answer metadata is important because it (a) uncovers patterns in developer struggles, (b) highlights gaps in existing support systems, and (c) provides actionable insights to prioritize improvements in XAI techniques. To evaluate the severity of each challenge, we compared the answer metadata of questions associated with each of the challenges. Such metadata includes the presence of answers (including acceptable ones) and the delay in receiving acceptable answers. For example, questions with a particular challenge that have a lower percentage of acceptable answers and a higher delay in receiving them are considered more severe.

**RQ3. What is the perceived severity of the XAI challenges based on practitioners' ability to use XAI techniques effectively in their work?** Understanding practitioners' perspectives on the severity of XAI challenges is crucial to improving their ability to use XAI techniques in real-world applications effectively. We surveyed 52 practitioners and asked them to rank the challenges by perceived severity when using XAI techniques.

**RQ4. What improvements or features can make XAI techniques more accessible and easier to use?** Understanding what improvements or features can make XAI techniques more accessible and easier to use is essential to bridge the gap between technical capabilities and practical adoption. Identifying these enhancements can help address usability barriers, wider adoption among practitioners and ensure XAI tools are effectively integrated into real-world workflows. We seek the agreement of practitioners on several proposed improvements and also gather their open opinions on additional features or changes that could enhance the accessibility and usability of XAI techniques.

This study identifies key challenges practitioners face with XAI techniques and their severity, offering actionable insights to develop more effective and user-focused solutions and serve as a foundation for future research.

**Replication Package** that includes the data to answer our RQs can be found in our online appendix [2].

## 2 RESEARCH METHODOLOGY

Figure 2 shows the schematic diagram of our exploratory study. First, we collected 663 SO questions related to XAI techniques and manually analyzed them to identify the challenges reported in these questions. Our analysis produced a challenge catalog. Second, we analyzed the correlation between each identified challenge and answer metadata (e.g., the presence of acceptable answers) to estimate the severity of each challenge. We then surveyed 52 practitioners. Participants assessed each challenge based on its impact on their ability to use XAI techniques effectively. Finally, we explored practical insights to improve the XAI accessibility and usability. The following sections discuss the key steps of our methodology.

### 2.1 Data Collection and Filtration

We collected the July 2024 data dump of SO from the Stack Exchange site [40], which was the most recent data dump available when we began our study. We then applied a tag-based filtration to extract the questions related to XAI techniques. We analyzed the entire SO question tag list and identified nine tags associated with XAI techniques. They are "LIME," "SHAP," "Breakdown", "PyExplainer", "Saliency Map", "anchors", "TimeLime", "LOCO", and "GradCAM". We found a total of 695 questions that contained at least one of these selected tags.

We then removed 21 duplicate questions from the 695 identified questions. In this study, we attempted to produce a catalog of XAI challenges. However, not all questions discuss such challenges, even though they have tags related to XAI techniques. Therefore, we initially screened each question and discarded 11 that did not discuss any challenge. After this screening, we found 663 questions that primarily focused on LIME and SHAP techniques. We verified and confirmed that there were no false positive samples in the dataset, and we thus proceeded with these 663 questions.

### 2.2 Challenge Catalog

We manually analyzed 663 questions and spent 160 person-hours to create an XAI challenge catalog. For qualitative analysis, we followed Cruzes et al.'s guidelines [5]. Cruzes et al.'s guidelines for qualitative analysis focus on systematically collecting and analyzing data by using techniques like thematic coding to identify patterns, triangulation to validate findings, and transparency to ensure reproducibility. These guidelines are relevant to us because they help ensure our analysis of the 663 XAI-related questions is rigorous, credible, and systematic, resulting in a reliable challenge catalog. This involved analyzing data to uncover and develop themes from a collection of questions, which is widely used in software engineering (SE) [3]. We involve two independent annotators to label the sampled data with XAI challenges. These annotators are: (a) the lead author of this paper, who has over 6 years of development experience and 4 years of research experience specializing in XAI, and (b) a skilled programmer with over 3 years of development experience, serving as a summer intern. They began by discussing a few questions and their possible labels to build an initial understanding. Then, each annotator independently analyzed and labeled an initial set of 50 questions. Their agreement level, measured using Cohen's Kappa [27], was 0.70. To resolve disagreements, they held discussions in the presence of the first co-author, who has over 10 years of development experience and 6 years of research experience with SO data. After these discussions, the annotators developed a strong shared common understanding. They then independently labeled the remaining questions and measured the agreement level again. This time, the agreement was 0.93, indicating perfect agreement. Any remaining disagreements were resolved with the help of the first co-author. This process resulted in a catalog of seven key XAI challenges that developers face.

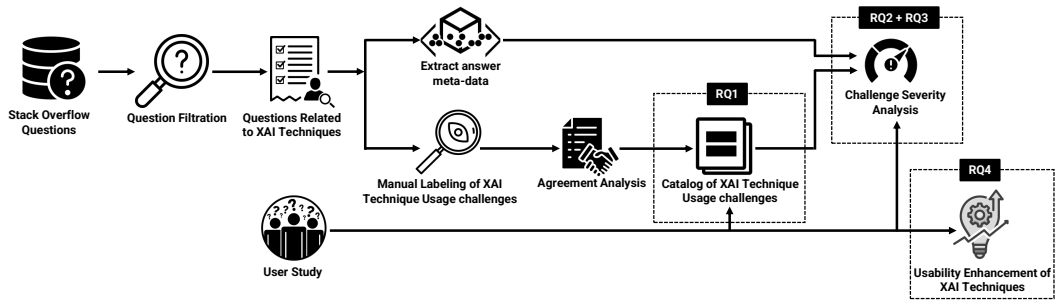

Figure 2: Schematic diagram of our exploratory study.

## 2.3 Challenge Severity (Correlation analysis)

To assess the severity of each XAI challenge developers face, we analyzed the correlation between challenge types and their associated answer metadata. Specifically, we recorded whether a question received an accepted answer, the time delay in receiving the accepted answer, and the total number of answers. The delay was calculated as the time difference (in minutes) between the submission of the accepted answer and the posting time of the question. These metrics were compared across challenge types to identify where developers face the most significant obstacles. For example, questions of a particular challenge with fewer accepted answers, longer delays, and a higher percentage of unanswered questions were classified as more severe. This analysis provides insights into critical challenges in XAI adoption and highlights areas for improvement.

## 2.4 User Study

We conducted an online survey to understand how frequently practitioners encounter our identified XAI challenges, including any additional ones, and to evaluate their severity. Additionally, we gathered participants' perspectives on improving XAI techniques to enhance usability. The survey design followed the guidelines for personal opinion surveys proposed by Kitchenham and Pfleeger [10]. We also adhered to the best practices and addressed ethical considerations outlined in prior works [8, 38]. Notably, the survey underwent revisions and received approval from our institutional Behavioral Research Ethics Board.

**Survey Design:** Our survey includes different types of questions (e.g., multiple-choice and free-text answers). Before asking questions, we explain the purpose of the survey and our research goals to the participants. We ensure the survey participants that the information they provide must be treated confidentially [21]. We first piloted the preliminary survey with six practitioners. We collect their feedback on (1) whether the length of the survey was appropriate and (2) the clarity and understandability of the terms. We then perform minor modifications to the survey draft based on the received feedback and produce a final version. We inform the estimated time (i.e., 10-15 minutes) required to complete the survey to the participants based on the pilot survey. We exclude the six responses from the pilot survey from the presented results in this paper. Our survey comprises five parts as follows.

**(i) Consent and Prerequisite:** In this part, we ask the participants to confirm whether they consent to participate in this survey. We also ask questions to confirm whether how much they familiar with XAI techniques and have experience in AI/ML, especially Python. Otherwise, we did not allow them to participate in our survey. In this section, participants were asked to confirm their consent to participate in the survey. We also verified their familiarity with XAI techniques and experience in AI/ML. Participants who had not used XAI techniques were not allowed to proceed.

**(ii) Participants Information:** In this part, we attempt to collect some information about the participants, such as years of experience in AI/ML, country of residence, their current profession, and which XAI techniques have used in their work.

**(iii) Severity Analysis:** In this section, we asked participants to rate the severity of each challenge based on its impact on their ability to effectively use XAI techniques in their work. Each challenge was presented with a 5-point Likert scale ranging from '*Not Severe*' to '*Extremely Severe*'. We also asked them to report the challenge that most affected their use of a specific XAI technique.

Table 1: Improvements/features for enhancing the usability of XAI techniques

| ID | Improvements/Features |
|---|---|
| BD | *Better Documentation and Tutorials:* Clear guides and easy-to-follow tutorials to help you learn and use XAI techniques. |
| IV | *Improved Visualization Tools:* Better tools for showing and understanding the results of XAI techniques |
| EI | *Enhanced Integration with Popular ML Frameworks:* Easier to use XAI techniques with popular machine learning tools like TensorFlow, PyTorch, and scikit-learn |
| CF | *More Customizable and Flexible Tools:* Tools that you can adjust and tailor to fit different needs |
| PO | *Performance Optimization:* Faster and more efficient XAI techniques that use fewer resources |
| UI | *User-Friendly Interfaces:* Simple and intuitive interfaces that make XAI techniques easier to use |
| CE | *Comprehensive Examples and Case Studies:* Practical examples and real-world scenarios to show how XAI techniques can be used |
| BS | *Better Support and Community Engagement:* More help and an active community for troubleshooting and sharing tips |
| AE | *Automated Explanation Generation:* Tools that automatically create explanations, saving you time |
| SC | *Simplified Configuration and Setup:* Easier setup and configuration so you can get started with XAI techniques quickly |
| CG | *Clearer Guidance on Best Practices:* Straightforward advice on the best ways to use XAI techniques effectively |
| EE | *Enhanced Error Handling and Debugging:* Better tools for finding and fixing problems with XAI techniques |

Table 2: Experience and profession of the participants (**MLE/C:** Machine Learning Engineer/Consultant, **AIR:** AI Researcher, **DAS:** Data Analyst/Scientist, **SE:** Software Engineer, **PM:** Product Manager, **AC:** Academician).

| Experience (Years) | | | | Profession | | | | | |
|---|---|---|---|---|---|---|---|---|---|
| ≤ 2 | 3-5 | 6-10 | ≥10 | MLE/C | AIR | DA/S | SE | PM | AC |
| 18 (34.62%) | 19 (36.54%) | 14 (26.92%) | 1 (1.92%) | 16 (30.77%) | 7 (13.46%) | 4 (7.69%) | 4 (7.69%) | 2 (3.85%) | 19 (36.54%) |

**(iv) Prevalence of XAI challenges:** In this part, we presented the XAI challenges identified through our qualitative analysis and asked participants to indicate which challenges they had encountered in their work. Additionally, we provided an option for participants to report any other challenges they faced.

**(v) Improvements/Features to Enhance Usability:** Finally, we asked participants for their opinions on improvements or features that could make XAI techniques more accessible and easier to use. In particular, we provided 12 improvement options, as shown in Table 1 and asked them to indicate their agreement using a 5-point Likert scale ranging from '*Strongly Agree*' to '*Strongly Disagree.*' However, there was an option to report additional improvements/features freely. Additionally, we invited participants to suggest the best way to improve one specific aspect of XAI tools to enhance their user-friendliness.

**Participants:** We recruit participants in the following two ways.

**(i) Snowball Approach:** We seek participants from universities/companies globally through personal interactions. We then used a snowballing strategy [4] to urge participants to refer our survey to colleagues with similar experiences and desire to join. In this approach, we found 30 participants. However, 17 of them did not meet our constraints. We thus finally allowed the remaining 13 to participate in our survey.

**(ii) Open Circular:** To find potential participants, we post a description of this study and our research goals in the specialized Facebook groups where professional software developers discuss their programming problems and share software development resources.

We also use LinkedIn as a research tool to reach potential participants because it is one of the largest professional social networks in the world. This approach identified 54 interested participants, resulting in 39 valid responses.

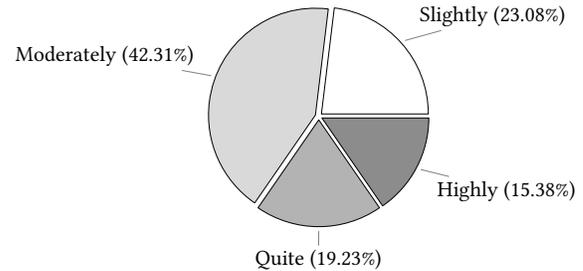

Figure 3: Familiarity with XAI techniques

Table 2 summarizes the participants' experience and profession. The majority of the participants were 'Academicians' (36. 54%) and 'Machine Learning Engineers or Consultants' (30. 77%). Over 50% of the participants had three or more years of AI/ML experience, lending credibility to the findings. Additionally, Figure 3 illustrates that approximately 77% of participants reported being highly, quite, or moderately familiar with XAI techniques. They participated in this survey from different countries globally, including the USA, Canada, France, Germany, Ireland, Russia, and Bangladesh. Such statistics ensure the reliability and diversity of the collected responses and our findings' validity.

## 3 STUDY FINDINGS

We collected 663 questions from SO that discussed XAI challenges and analyzed them to produce a challenge catalog. We ask four research questions in this study and answer them carefully with the help of our empirical and qualitative findings and insights from practitioners' perspectives.

### Table 3: Challenges of XAI techniques

| ID | Identified Challenges |
|---|---|
| CI | Compatibility Issues |
| ID | Installation Difficulties |
| DT | Data Transformation / Integration Issues |
| VP | Visualization and Plotting Issues |
| MI | Model Integration Issues |
| PI | Performance and Resource Issues |
| DI | Disagreement Issues |

### 3.1 Answering RQ1: Production of an XAI Challenge Catalog and Challenge Prevalence

Our manual investigation that analyzed 663 SO questions produced a catalog of *seven* XAI challenges, as shown in Table 3. The challenges are discussed below.

**(C1) Compatibility Issues** arise when XAI techniques fail to function correctly across different ML frameworks, libraries, or hardware due to mismatched dependencies, conflicting versions, or unsupported configurations. For example, SHAP fails to generate explanations for PyTorch-based RNNs, preventing model interpretation. Broadcasting issues with Ndarrays cause runtime errors due to incompatible shapes, while tensor size mismatches lead to shape errors that disrupt computations. Non-callable models arise due to incorrect object types or framework constraints, limiting their usability. SHAP DeepExplainer encounters failures when handling models with multiple separate inputs, resulting in missing or inconsistent attributions. Similarly, Keras-LSTM models face SHAP failures that obstruct meaningful explanations. Python LIME suffers from indexing issues, particularly with high-dimensional data, and Shapley value dimension mismatches lead to incorrect or unusable results. Consider the motivational example in Section 1 and Figure 1, which highlights a compatibility issue.

**(C2) Installation and Package Dependency Issues** hinder the setup of XAI tools and frameworks due to dependency conflicts or insufficient documentation. According to our analysis, common issues include SHAP installation failures in Jupyter Notebook, package import errors, permission-denied issues with library functions, kernel crashes when executing SHAP, and installation errors in PyCharm. CI/CD pipelines on GitLab also encounter SHAP installation failures, while conflicts with pandas and NumPy cause functionality issues after installation. Figure 4 presents an SO question where a user faces difficulties installing SHAP. Such issues slow down the adoption of explainability techniques and reduce the transparency and interpretability of ML systems.

**(C3) Visualization and Plotting Issues** affect the clarity and interpretability of XAI outputs due to unclear feature attributions,

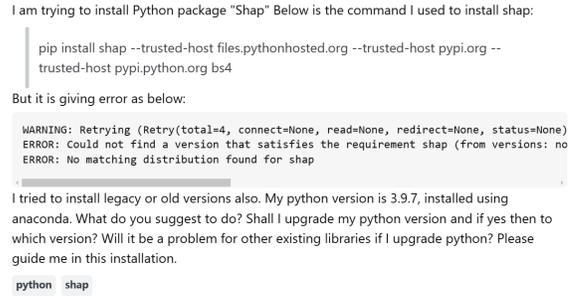

Figure 4: An example of a question on SO related to XAI Model Installation Issues [26]

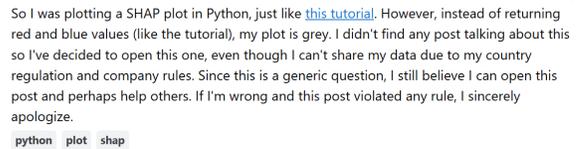

Figure 5: An example of a question on SO related to XAI Visualization Discrepancy [23]

misleading graphs, or compatibility issues with visualization libraries. Our analysis reveals some issues that include difficulties in plotting LIME with many features, generating SHAP summary plots for multi-label classification, creating SHAP scatter plots for LightGBM, and handling inconsistencies in SHAP summary and bar plots. Additional issues involve shap.dependence_plot() triggers index errors, LIME visualizations for BERT transformers lead to memory errors, and missing emoji features in LIME figures reduce interpretability. Figure 5 illustrates a SHAP visualization issue where expected red-blue feature attributions appear entirely gray. This reduces interactivity and interpretability in XAI outputs and emphasizes the need for robust visualization support in tools like SHAP. Inconsistencies in plots, discrepancies in visual outputs, and difficulties understanding feature importance often hinder developers' ability to extract meaningful insights [12, 17].

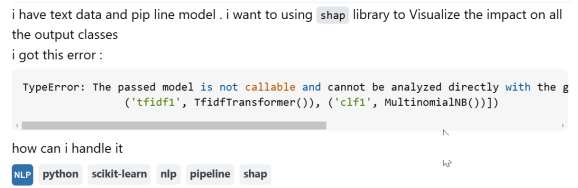

Figure 6: An example of a question on SO related to XAI Model Integration Issues [19]

**(C4) Model Integration Issues** occur when embedding XAI techniques into ML pipelines or applications, where API constraints,

Figure 7: An example of a question on SO related to Data Transformation or Integration [32]

computational overhead, or model-specific limitations hinder their practical deployment. Unlike compatibility issues, which stem from framework conflicts, integration challenges occur even when XAI methods function but fail to align with a model's structure or workflow. These issues affect complex ML systems like ensemble models and neural networks and limit their adoption in applications such as review recommendations and autonomous systems [41]. Commonly reported issues on SO include transforming SHAP values for LightGBM Tweedie objectives, applying LIME to time series classification, and integrating SHAP with custom ElasticNet models in R. Figure 6 illustrates a case where a user struggles to integrate SHAP into a text classification task, facing errors or misaligned outputs. Such challenges reduce the usability of XAI techniques, often requiring custom adaptations to fit into existing workflows.

**(C5) Data Transformation/Integration Issues** arise when converting, formatting, or merging data from different sources for use in XAI models. These challenges lead to inconsistencies, loss of information, or reduced interpretability and make it difficult to generate reliable explanations. Common issues include formatting LIME for DeepSHAP, merging SHAP values across class labels into a single matrix, and interpreting SHAP in autoencoders. Another challenge involves transforming SHAP values back into model coefficients while preserving interpretability. Figure 7 illustrates an issue in multi-class classification, where a user struggles to combine SHAP values from different classes into a unified representation. Since SHAP computes values separately for each class, aggregation becomes complex, especially when feature importance varies across classes. The lack of a clear merging mechanism makes it difficult to balance or prioritize SHAP contributions effectively.

**(C6) Performance and Resource Issues** impact the efficiency of XAI models due to high computational costs, memory consumption, and extended processing times, limiting scalability and practical use. Our analysis identifies common issues such as SHAP values for invalid features, difficulties in extracting key features, missing feature scores in LIME for positive class explanations, LIME prediction probability mismatches, and SHAP statistical inconsistencies. Additional cases include Databricks HTML output errors, Keras model evaluation failures, and performance bottlenecks in SHAP's

Figure 8: An example of a question on SO related to Performance and resource Issues [31]

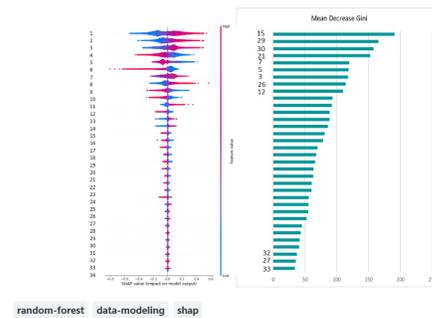

Figure 9: An example of a question on SO related to Disagreement problems [18]

DeepExplainer. Figure 8 highlights a challenge in using SHAP for local accuracy evaluation. While designed to quantify individual feature contributions, SHAP sometimes produces inconsistent results due to its reliance on game-theoretic principles, which may not fully align with nonlinear models. Improper data preprocessing (e.g., scaling or encoding) and high computational demands further complicate its use. These challenges underscore the need for optimized algorithms and clearer documentation to enhance SHAP's reliability in performance-critical environments.

**(C7) Disagreement Issues** arise when different XAI methods generate conflicting explanations that lead to feature attributions and interpretability inconsistencies. Common cases include discrepancies in SHAP explainer bar charts, differences in SHAP instance attributions, variable ranking mismatches, and inconsistent SHAP explanations. Figure 9 illustrates a disagreement between XAI techniques like LIME and SHAP when applied to the same ML model. These methods often produce conflicting feature importance rankings or interpretations for identical datasets and thus make it difficult for developers to determine which explanation is more reliable. Such inconsistencies complicate decision-making and reduce trust in XAI outputs.

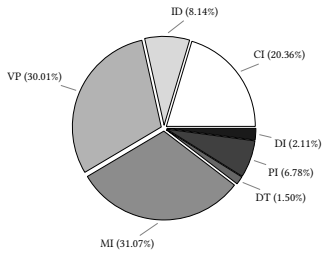

Figure 10: Ratio of each challenge

Figure 10 illustrates the prevalence of key challenges (manual analysis) in adopting XAI techniques. Model Integration Issues (31.07%) and Visualization and Plotting Issues (30.01%) are the most reported, highlighting difficulties in embedding XAI tools into ML workflows and interpreting their outputs. SHAP and LIME often struggle with complex architectures [7], while visualization inconsistencies hinder trust in explanations [12, 17]. Compatibility Issues (20.36%) further complicate adoption, with framework mismatches preventing seamless execution [1, 13]. Installation Difficulties (8.14%) create friction due to dependency conflicts and undocumented setup requirements [26, 29, 36, 37], while Performance Issues (6.78%) reflect concerns over high computational costs, especially for large datasets [46]. Less frequently reported in SO but still critical, Disagreement Issues (2.11%) expose inconsistencies in feature attributions across XAI methods, making explanations unreliable [17]. Data Transformation Issues (1.50%) indicate challenges in preparing data for explainability, particularly in feature encoding and scaling [41]. The dominance of integration and visualization issues underscores the need for more adaptable, efficient, and standardized XAI frameworks to enhance usability and trust.

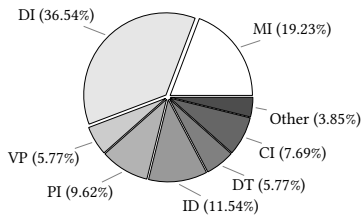

Figure 11: Challenges mostly affect when using XAI

Figure 11 highlights the challenges practitioners face (user study) when using XAI. Disagreement Issues (36.54%) rank highest, which indicates that conflicting explanations from tools like LIME and SHAP create significant frustration and distrust. In contrast, earlier analysis reported Disagreement Issues as the least frequent (2.11%). This discrepancy suggests that users often focus first on resolving technical barriers, but once they integrate XAI tools, inconsistencies in explanations become the primary concern, affecting decision-making and reliability. Model Integration Issues (19.23%) remain a key concern, which indicates that integration problems occur more often when using XAI techniques. Installation Difficulties (11.54%), Compatibility Issues (7.69%), and Performance Issues (9.62%) persist but cause less disruption than inconsistent explanations. Visualization (5.77%) and Data Transformation Issues (5.77%) rank lowest, likely because these challenges, though technical, do not fundamentally hinder usability. The contrast between reported issue frequency and perceived impact underscores a critical insight: technical barriers are common, but usability challenges—especially explanation inconsistencies—pose the greatest obstacle to trust and adoption. Addressing integration issues and feature attributions' inconsistencies remains essential for improving XAI's practical use.

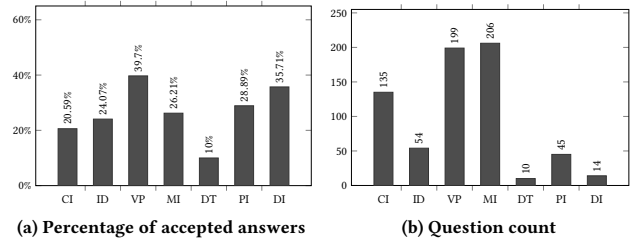

Figure 12: XAI challenges vs. presence of accepted answers

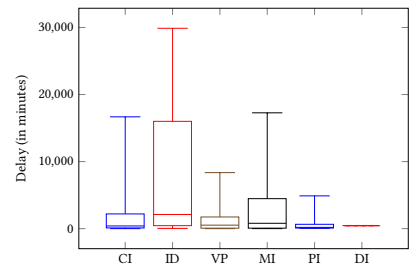

Figure 13: Time delay in receiving acceptable answers

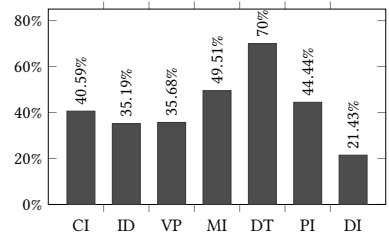

Figure 14: Percentage of unanswered questions

## 3.2 Answering RQ2: Answer Metadata-Based Challenge Severity Estimation

In this section, we analyzed the correlation between questions of the seven XAI challenges and the corresponding answer metadata. Figure 12 shows the percentage (Figure 12a) and count (Figure 12b) of *resolved* (i.e., received acceptable answers) questions from each challenge category. Overall, the acceptance rate remains low across all categories, indicating persistent challenges in finding definitive solutions for XAI-related issues.

Visualization Issues (39.7%) has the highest percentage of accepted answers. This suggests that while these challenges occur frequently—199 visualization-related questions out of 663—solutions are available and applicable. Disagreement Issues received 35.71% of accepted answers, although the question count was low (14). Challenges such as Installation Difficulties, Model Integration, and Performance Issues show an acceptance rate between 24% and 29%, indicating that while solutions exist, they may require significant effort to implement. Compatibility Issues (20.59%) have a lower acceptance rate that reflects the difficulty of resolving framework-specific constraints and dependency conflicts. Data Transformation Issues (10%) have the lowest acceptance rate, though their low question count (only 10) suggests that fewer developers report such problems, possibly because they arise in specialized use cases. The low resolution rate for compatibility and data transformation issues indicates that existing solutions may be insufficient or poorly documented, leaving many questions unresolved.

As shown in Figure 13, the overall delay in getting accepted answers is high. The median delay in accepted answers highlights the severity of Installation and Package Dependency Issues (2114 minutes), which also have a low acceptance rate (24.14%). Resolving installation challenges, such as dependency conflicts and version mismatches, proves both difficult and time-consuming, often requiring extensive trial and error. Model Integration (800 minutes) and Visualization (523 minutes) also experience significant delays, indicating the complexity of embedding XAI techniques into diverse ML architectures and resolving visualization inconsistencies. Integration issues often require modifying APIs and adapting models. Compatibility Issues (403 minutes) and Disagreement Issues (451 minutes) have moderate delays, reflecting ongoing difficulties but slightly faster resolution rates. Performance Issues (189 minutes) have the shortest median delay. The trend indicates that installation, integration, and visualization issues demand the most time to resolve, emphasizing the need for improved documentation and automated support to streamline troubleshooting.

Figure 14 shows that Data Transformation (70%), Model Integration (49.51%), and Performance Issues (44.44%) have the highest unanswered rates. Data Transformation Issues, with both the highest unanswered rate (70%) and lowest accepted answer rate (10%), suggest a lack of clear solutions. Model Integration, Compatibility (40.59%), and Visualization Issues (35.68%) also have high unanswered rates, showing ongoing challenges in applying XAI techniques. Disagreement Issues (21.43%) have the lowest unanswered rate, although fewer questions are reported. While some challenges receive responses, complex issues like Data Transformation and Model Integration remain unanswered.

### 3.3 Answering RQ3: Challenge Severity Perception by Practitioners

Figure 15 shows that Disagreement Issues (42.31%) are the most severe, despite a moderate accepted answer rate (35.71%) and the lowest unanswered rate (21.43%). This suggests that while solutions exist, inconsistencies between explainers like SHAP and LIME continue to frustrate users, reducing trust in XAI. Model Integration (34.61%) is also highly severe, with a high unanswered rate (49.51%) and long delays (800 minutes), reflecting difficulties in embedding XAI tools into ML workflows. Performance Issues (42.31% highly severe) highlight concerns about computational overhead. Compatibility Issues were assessed as moderately severe by 50% of the practitioners. Visualization Issues (40.38% moderately severe) remain common but have a higher resolution rate (39.7% accepted answers). However, Not Severe and Mildly Severe ratings are minimal across most of the challenges, indicating that practitioners generally perceive XAI issues as significant obstacles. Installation (25%) Issues have the highest proportion of Not Severe, and Visualization Issues are highest in Midly Severe, suggesting that while these challenges exist, they are manageable. Overall, Disagreement Issues persist despite available solutions that lead to user frustration. Model Integration, Data Transformation, Compatibility, and Performance Issues are mostly moderately severe, requiring better integration strategies and computational optimizations to enhance XAI adoption.

### 3.4 Answering RQ4: Improvement or Feature Recommendations by Practitioners

Figure 16 shows the agreement with proposed improvements to enhance the accessibility and usability of XAI techniques. Better documentation (55.77% strongly agree) receives the highest support, emphasizing the need for clearer guides and structured tutorials. XAI tools often suffer from inconsistent or scattered documentation that make adoption difficult for practitioners. Similarly, Clearer Guidance on Best Practices (48.07%) and Simplified Configuration (40.38%) receive strong agreement, reflecting the demand for standardized methodologies and easier setup to reduce onboarding complexity. User-friendly interfaces (38.46% strongly agree, 28.85% agree) and Improved Visualization Tools (32.69% agree) highlight usability as a key concern. Many XAI techniques require technical expertise, and more intuitive tools can lower the entry barrier. Enhanced Integration with ML Frameworks (30.77% strongly agree, 28.85% agree) and Performance Optimization (34.62% strongly agree) also gain significant support. Improvement options with neutral to lower agreement include Automated Explanation Generation (30.77% neutral) and More Customizable Tools (34.62% neutral), suggesting that while useful, these features are not as urgent. Overall, clearer documentation, standardized best practices, and streamlined integration demand urgent improvements for XAI adoption. While enhanced visualization and automation are valued, foundational usability challenges take priority. We requested additional recommendations to enhance XAI techniques, but practitioners did not provide any significant suggestions.

## 4 DISCUSSION

### 4.1 How Difficult Are XAI Techniques?

Figure 17 illustrates the perceived complexity of using XAI techniques. A majority of practitioners (46.15%) find XAI either very difficult or difficult, while 46.15% consider it somewhat difficult. Only 7.69% rate XAI as easy, showing that most users face significant challenges. The high difficulty ratings align with earlier findings on installation issues, inconsistencies in explanations, and integration challenges. Many practitioners struggle with complex setups, unclear documentation, and framework compatibility, making XAI tools harder to adopt. Additionally, the lack of standardized best practices and intuitive visualizations further contributes to the learning curve. These results highlight the need for better documentation, more intuitive interfaces, and streamlined integration to improve XAI usability. While XAI provides valuable insights, its complexity remains a major barrier, requiring focused improvements to enhance accessibility and adoption.

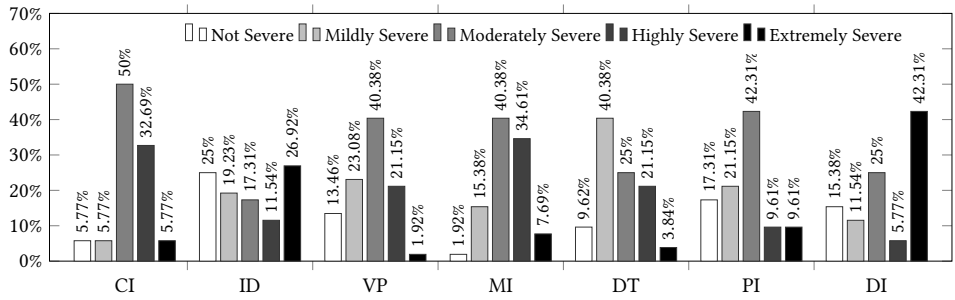

Figure 15: The severity of each challenge (**CI** = Compatibility Issues, **ID** = Installation Difficulties, **VP** = Visualization and Plotting Issues, **MI** = Model Integration Issues, **DT** = Data Transformation / Integration Issues, **PI** = Performance and Resource Issues, **DI** = Disagreement Issues).

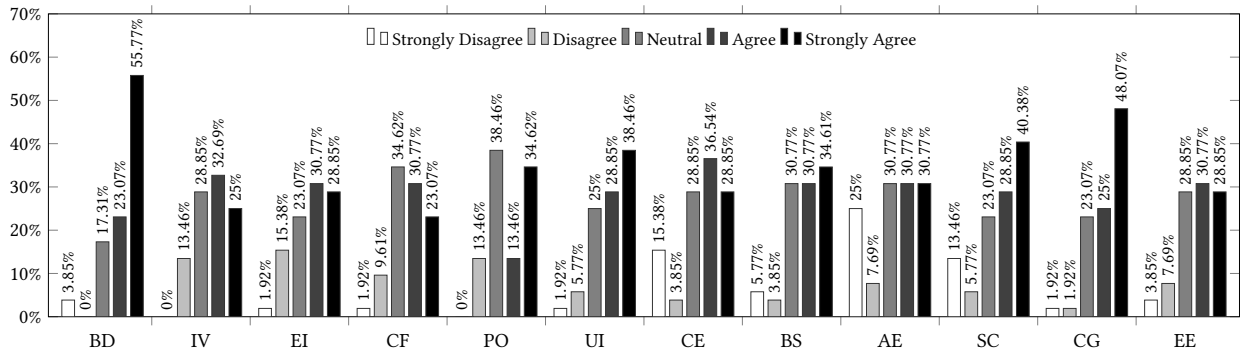

Figure 16: Improvements or feature recommendations to make XAI techniques more accessible and easier to use (**BD**: Better Documentation and Tutorials, **IV**: Improved Visualization Tools, **EI**: Enhanced Integration with Popular ML Frameworks, **CF**: Customizable and Flexible Tools, **PO**: Performance Optimization, **UI**: User-Friendly Interface, **CE**: Comprehensive Examples and Case Studies, **BS**: Better Support and Community Engagement, **AE**: Automated Explanation Generation, **SC**: Simplified Configuration and Setup, **CG**: Clearer Guidance on Best Practices, **EE**: Enhanced Error Handling and Debugging)

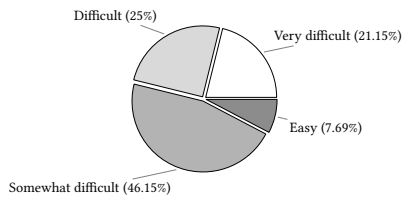

Figure 17: Complexity of using XAI techniques

## 4.2 If You Could Change One Aspect of XAI Tools to Make Them More User-Friendly, What Would It Be?

We asked practitioners to mention one aspect of XAI tools to make them more user-friendly. The responses highlight key areas for improvement, summarized below (full recommendations can be found in our online appendix [2]).

**Ease of Use & Installation:** 23.1% of practitioners seek simpler setup, plug-and-play functionality, and fewer dependency issues. One respondent noted, "Installation should be easier along with consistent explanations," highlighting frustrations with complex setup and inconsistencies across environments.

**Improved Visualization & Interpretability:** 17.3% of practitioners want clearer, interactive, and customizable visualizations. A respondent suggested, "Layered explanations should start simple and allow deeper exploration," emphasizing the need for adaptable visualizations suited to different expertise levels.

**Addressing Disagreement & Consistency Issues:** 15.4% of practitioners emphasize the need for handling inconsistencies across XAI methods. One user stated, "XAI tools should acknowledge disagreements and guide users in selecting reliable explanations," pointing to confusion caused by conflicting outputs from methods like SHAP and LIME.

**Model-Agnostic & Flexible Integration:** 13.5% of practitioners prefer framework-independent XAI tools that integrate easily with TensorFlow, PyTorch, and other ML frameworks. A respondent shared, "A tool where users can integrate models and receive error-specific explanations would improve usability," showing the need for adaptable XAI solutions.
**User-Friendly Interfaces & Accessibility:** 11.5% of practitioners want intuitive UI, interactive dashboards, and simplified tools for non-experts. One user suggested, "An open-source UI for testing and exploring explanations would be a game-changer," reflecting the demand for more accessible interfaces.

These findings highlight installation complexity, inconsistencies in explanations, visualization gaps, and integration difficulties as key challenges. Addressing them will improve XAI usability, adoption, and trust among practitioners.

### 4.3 How Can XAI Designers and Researchers Improve Usability?

Our findings highlight key areas for XAI designers and researchers to enhance usability and adoption. Disagreement Issues (42.31% extremely severe) demand standardized explainability metrics to reduce inconsistencies between methods like SHAP and LIME. Model Integration Challenges (34.61% highly severe, 49.51% unanswered) call for plug-and-play integration with TensorFlow and PyTorch. Performance Issues (42.31% severely severe or worse) necessitate lightweight, optimized techniques for large-scale models. Data Transformation (70% unanswered) requires automated feature preprocessing pipelines to ease adaptation. Visualization Challenges (40.38% moderately severe) suggest the need for layered, interactive explanations for better interpretability. The complexity of using XAI (46.15% somewhat difficult, 46.15% difficult/very difficult) highlights the need for clearer selection guidelines and best practices. Installation difficulties and high unanswered rates point to the need for better dependency management, pre-configured environments, and interactive documentation. Finally, 38.46% of practitioners strongly agree on the need for intuitive interfaces, reinforcing the importance of user-friendly workflows and interactive dashboards. Addressing these issues through standardization, automation, and better UI design will significantly improve XAI accessibility, trust, and real-world adoption.

## 5 THREATS TO VALIDITY

**External validity** concerns the generalizability of our findings. Our analysis is based on challenges reported in SO questions, which may not cover all XAI-related issues. However, we mitigated this limitation by analyzing all XAI-related questions and surveying 52 practitioners to validate our findings. While this strengthens reliability, we caution against overgeneralization. **Internal validity** relates to potential biases in data selection. We identified SO tags for XAI techniques and extracted relevant questions, but tag limitations may exclude some target questions or introduce false positives. To address this, we manually validated each selected question, ensuring no false positive samples.

**Annotation bias** is a concern when labeling the XAI challenge catalog. Two annotators performed the labeling, achieving near-perfect agreement ($\kappa = 0.93$). Any disagreements were resolved with an expert, reducing subjective bias. **Sampling bias** may arise from snowball sampling, as referrals can limit diversity. To minimize this, we used an open circular approach and collected responses anonymously. As shown in Table 2, our participants have diverse backgrounds, enhancing the validity and applicability of our survey. The sample size of 52 participants further mitigates individual biases in responses.

## 6 RELATED WORK

Explainable AI (XAI) research has primarily focused on improving model interpretability [6], yet practical adoption challenges remain underexplored, particularly those surfaced in real-world developer discussions on platforms like SO. Studies highlight explainability inconsistencies [35], usability concerns [14], and integration difficulties, but few quantify how these issues affect developers or how they navigate them in practice.

Early work on post hoc explanation methods such as SHAP and LIME [16, 30] improved transparency but introduced inconsistencies across explainers, leading to uncertainty in decision-making [11, 35]. While studies discuss performance-interpretability trade-offs [9, 15, 34], they often overlook practical barriers like compatibility, computational overhead, and workflow integration. Research on AI usability [14] calls for tools that align with developer needs, yet little examines how XAI challenges manifest in real-world software development. Existing work on developer struggles in SE platforms [20, 22, 33, 44] does not specifically address XAI adoption barriers, and studies on community-driven support [39, 42, 45] focus on discussion patterns rather than usability challenges.

We bridge this gap by analyzing SO discussions to categorize key XAI challenges, validating findings with practitioners, and quantifying severity using answer metadata. Our study offers an empirical, user-driven perspective on XAI adoption barriers, providing actionable insights to improve usability, reliability, and real-world integration of XAI tools.

## 7 CONCLUSION

This study analyzed challenges in XAI adoption reported in 663 SO questions. Our findings reveal that model integration, visualization, and compatibility issues are the most reported technical barriers, while disagreement among explainers and performance constraints pose major usability concerns. Disagreement and model integration challenges have lower answer acceptance rates, long resolution times, and high unanswered percentages, indicating persistent difficulties. Despite community engagement, gaps remain in integration, explainability consistency, and performance optimization. We surveyed 52 practitioners. They highlighted the need for better documentation, enhanced visualization, seamless ML framework integration, and performance improvements. Addressing these requires standardized explainability frameworks, improved tool interoperability, and automated explanation generation to make XAI more transparent, reliable, and practical.

## ACKNOWLEDGMENTS

The first two authors contributed equally to this study. This research is supported in part by the Natural Sciences and Engineering Research Council of Canada (NSERC) Discovery Grants program, the Canada Foundation for Innovation's John R. Evans Leaders Fund (CFI-JELF), and by the industry-stream NSERC CREATE in Software Analytics Research (SOAR).